\documentclass[traditabstract]{aa}
\usepackage{txfonts}
\usepackage{epsfig}
\usepackage{lscape}
\bibpunct{(}{)}{;}{a}{}{,}    
\usepackage{graphicx,graphics}
\usepackage{color}
\usepackage{ulem}
\usepackage{longtable}
\usepackage{amsmath}
\usepackage{amssymb}
\usepackage{adjustbox}
\usepackage[breaklinks=true, colorlinks=true, urlcolor=blue, citecolor=blue, linkcolor=blue]{hyperref}

\begin{document}

\title{First direct electron temperature measurement in [\ion{O}{II}] zone in I~Zw~18}

\author{
        I.~A.~Zinchenko\inst{\ref{ARI},\ref{MAO}} \and
        J.~M.~V\'{i}lchez\inst{\ref{IAA}} \and
        C.~Kehrig\inst{\ref{IAA},\ref{newinst}} \and
        P.~Papaderos\inst{\ref{CAUP}} \and
        J.~E.~M\'{e}ndez-Delgado\inst{\ref{IAM}}
}
       
\institute{Astronomisches Rechen-Institut, Zentrum f\"{u}r Astronomie der Universit\"{a}t Heidelberg, M\"{o}nchhofstra{\ss}e 12-14, D-69120 Heidelberg, Germany\label{ARI}
\and
Main Astronomical Observatory, National Academy of Sciences of Ukraine, 
27 Akademika Zabolotnoho St., 03143, Kyiv, Ukraine\label{MAO}
\and
Instituto de Astrof\'{i}sica de Andaluc\'{i}a (CSIC), Apartado 3004, 18080 Granada, Spain \label{IAA}
\and
Observat\'orio Nacional/MCTIC, R. Gen. Jos\'e Cristino, 77, 20921-400, Rio de Janeiro, Brazil \label{newinst}
\and
Instituto de Astrofísica e Ciências do Espaço, Universidade do Porto – CAUP, Rua das Estrelas, PT4150-762, Porto, Portugal \label{CAUP}
\and
Instituto de Astronom\'{i}a, Universidad Nacional Aut\'{o}noma de M\'{e}xico, Ap. 70-264, 04510 CDMX, Mexico \label{IAM}
}

\abstract{
We present new precise measurements of electron temperatures and oxygen abundances in the southeast knot of I Zw 18, one of the most metal-poor blue compact dwarf galaxies known, using spectroscopic data from the Dark Energy Spectroscopic Instrument Data Release 1 (DESI DR1). For the first time in I Zw 18, we directly measure electron temperature in the low ionization zone using the rarely detected [\ion{O}{II}]$\lambda\lambda$7320,7330 doublet. We also detected [\ion{O}{III}]$\lambda$4363 and [\ion{S}{III}]$\lambda$6312 auroral lines, associated with high and intermediate ionization zones, respectively.
We derive T$_e$([\ion{O}{III}]) = 21\,200$\pm$860~K, T$_e$([\ion{O}{II}]) = 16\,170$\pm$950~K, and T$_e$([\ion{S}{III}]) = 17\,290$\pm$1750, highlighting a significant temperature difference between ionization zones. Using these direct temperature measurements, we determine a total oxygen abundance of 12+$\log$(O/H) = 7.066$\pm$0.046, $\log$(N/O) = -1.509$\pm$0.097, and $\log$(S/O) = -1.558$\pm$0.041. Our results extend the calibration of \( t_2-t_3 \) relations to the highest temperatures, providing important anchor points for the temperature structure of extremely metal-poor \ion{H}{II} regions, including high-redshift galaxies where direct temperature measurements are especially challenging.
}


\titlerunning{Direct electron temperature measurements in I Zw 18}
\authorrunning{Zinchenko et~al.}
\maketitle

\section{Introduction}

Galaxies with extremely low metallicity are rare in the nearby Universe. Nevertheless, they provide exceptional opportunities to study the physical conditions and processes that occurred in the early Universe and during the epoch of reionization. Therefore, research on these galaxies is crucial for enhancing our knowledge of the initial phases of galaxy evolution.

Chemical abundances in \ion{H}{II} regions of galaxies are most reliably determined using the direct method (also known as the T$_e$ method). This method relies on the direct measurement of electron temperature (T$_e$) using temperature-sensitive auroral lines, such as [\ion{O}{III}]$\lambda$4363, [\ion{N}{II}]$\lambda$5755, [\ion{S}{III}]$\lambda$6312,  and [\ion{O}{II}]$\lambda\lambda$7320,7330. These auroral lines are very weak and require deep, high-quality spectroscopy with large telescopes to be detected. Once the T$_e$ is determined from the ratio of auroral to nebular lines, and the electron density (n$_e$) is measured using density-sensitive line ratios, ionic abundances can be computed directly from the observed line intensities using statistical equilibrium equations~\citep[see, e.g.,][]{Osterbrock2006,PerezMontero2017}. 
For oxygen, often used as a gas-phase metallicity tracer, the total abundance is given by O/H = (O$^+$/H$^+$) + (O$^{++}$/H$^+$), where O$^+$ is derived from [\ion{O}{II}]$\lambda\lambda$3727,3729 lines and O$^{++}$ from [\ion{O}{III}]$\lambda\lambda$4959,5007 lines. For elements like nitrogen and sulfur, ionization correction factors (ICFs) are essential since N$^{++}$ and S$^{3+}$ cannot be observed in optical spectra, requiring correction by ICF.

In low-metallicity \ion{H}{II} regions, it is very common that only [\ion{O}{III}]$\lambda$4363 auroral line can be measured, providing estimation of T$_e$([\ion{O}{III}]) and, therefore, O$^{++}$/H$^+$, in O$^{++}$ zone. In this case, O$^+$/H$^+$ in low ionization O$^{+}$ zone is calculated assuming a relation between T$_e$([\ion{O}{II}]) and T$_e$([\ion{O}{III}]). A number of popular linear theoretical relations between T$_e$ in low and high ionization zones, also referred as \( t_2-t_3 \) relations, based on photoionization models have been proposed by \citet{Stasinska1980,Pagel1992,Garnett1992} among others, while others propose theoretical~\citep{PerezMontero2003, Izotov2006} or empirical~\citep{MendezDelgado2023} non-linear relations with saturation of the T$_e$([\ion{O}{II}]) at high T$_e$([\ion{O}{III}]) or even more complex functional relations \citep{ArellanoCordova2020,RickardsVaught2025}. Because [\ion{O}{II}] and [\ion{N}{II}] arise from similar low-ionization zones, it is frequently assumed that T$_e$([\ion{O}{II}])~$\approx$~T$_e$([\ion{N}{II}]) \citep[e.g.][]{Izotov2006,Croxall2016,Yates2020,Zurita2021}. Therefore, T$_e$([\ion{N}{II}]) is often used in \( t_2-t_3 \) relations, instead of T$_e$([\ion{O}{II}]).

Recent JWST/NIRSpec observations of star-forming galaxies at \(z \sim 2{-}3\) reveal that the T$_e$([\ion{O}{II}])--T$_e$([\ion{O}{III}]) relation at high redshift exhibits a shallower slope than that found for nearby galaxies, though it remains consistent once extremely metal poor (XMP) systems are included \citep{Cataldi2025}. Overall, the high-temperature regime is sparsely sampled observationally and therefore relies largely on model-based relations.
As such, local XMPs provide an important local benchmark to test the universality of this relation across cosmic time.

I~Zw~18 is a blue compact dwarf (BCD) galaxy, which represents one of the most extreme low-metallicity star-forming systems. With an oxygen abundance of approximately 3\% of the solar value, corresponding to 12+$\log$(O/H) $\sim$ 7.1–7.2 \citep[see, e.g.,][]{Izotov1998,Vilchez1998,Kehrig2016}, this galaxy serves as a critical laboratory for the examination of physical conditions and processes in the primordial universe and during the very first cycles of star formation. The galaxy exhibits active star formation with multiple stellar populations spanning ages from very young (< 30 Myr) to intermediate ages (100-800 Myr), providing insights into chemical evolution across different star formation episodes.

In this work, our aim is to obtain the first direct estimation of the electron temperature in the ionization zone of O$^+$ and the oxygen abundance in I Zw 18 using a new spectrum of its southeast (SE) star-forming (SF) knot published in the Dark Energy Spectroscopic Instrument Data Release 1~\citep[DESI DR1;][]{DESIDR1}.

\begin{table*}
\caption{Electron densities, [\ion{O}{III}] temperatures t$_3$, and element abundances in the I~ZW~18 SE knot}    
\label{table:abundance}      
\centering                          
\begin{adjustbox}{width=\linewidth}
\begin{tabular}{l c c c c c c c c c}        
\hline\hline                 
 Target ID  & n$_e$([\ion{S}{II}]) &  n$_e$([\ion{O}{II}])  & T$_e$([\ion{O}{III}]) & T$_e$([\ion{O}{II}]) & T$_e$([\ion{S}{III}]) & 12+$\log$(O$^{++}$/H) & 12+$\log$(O$^{+}$/H) & 12+$\log$(O/H) &  \\  
            & cm$^{-3}$   &  cm$^{-3}$  &K        & K         & K       &  dex       & dex            & dex             &            \\
\hline                        
39633324993414901  & <100 & 27$\pm$32 & 21200$\pm$860 & 16170$\pm$950 & 17290$\pm$1750 & 6.924$\pm$0.034 & 6.510$\pm$0.075 & 7.066$\pm$0.046 &  \\
\hline                                   
\end{tabular}
\end{adjustbox}
\end{table*}

\begin{figure}
\resizebox{0.90\hsize}{!}{\includegraphics[angle=000]{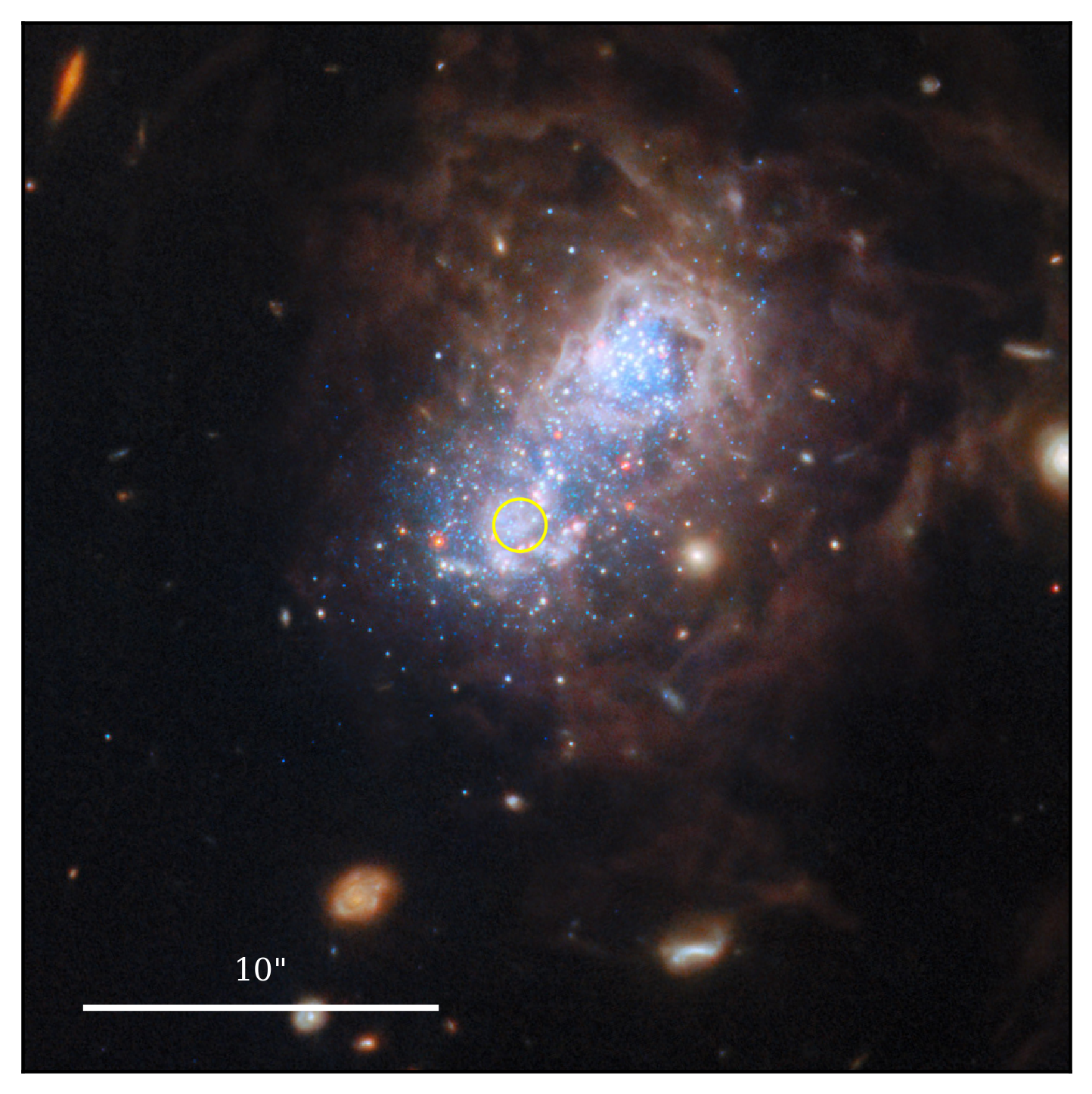}}
\caption{%
     Color composite JWST image of I Zw 18 made using four near-IR NIRCam bands (F115W, F200W, F356W, and F444W) by ESA/Webb, NASA, CSA \citep{Hirschauer2024}. North is up and east is left. Yellow circle represents DESI spectroscopic 1.5" aperture.
}
\label{figure:image}
\end{figure}

\begin{figure}
\resizebox{1.0\hsize}{!}{\includegraphics[angle=000]{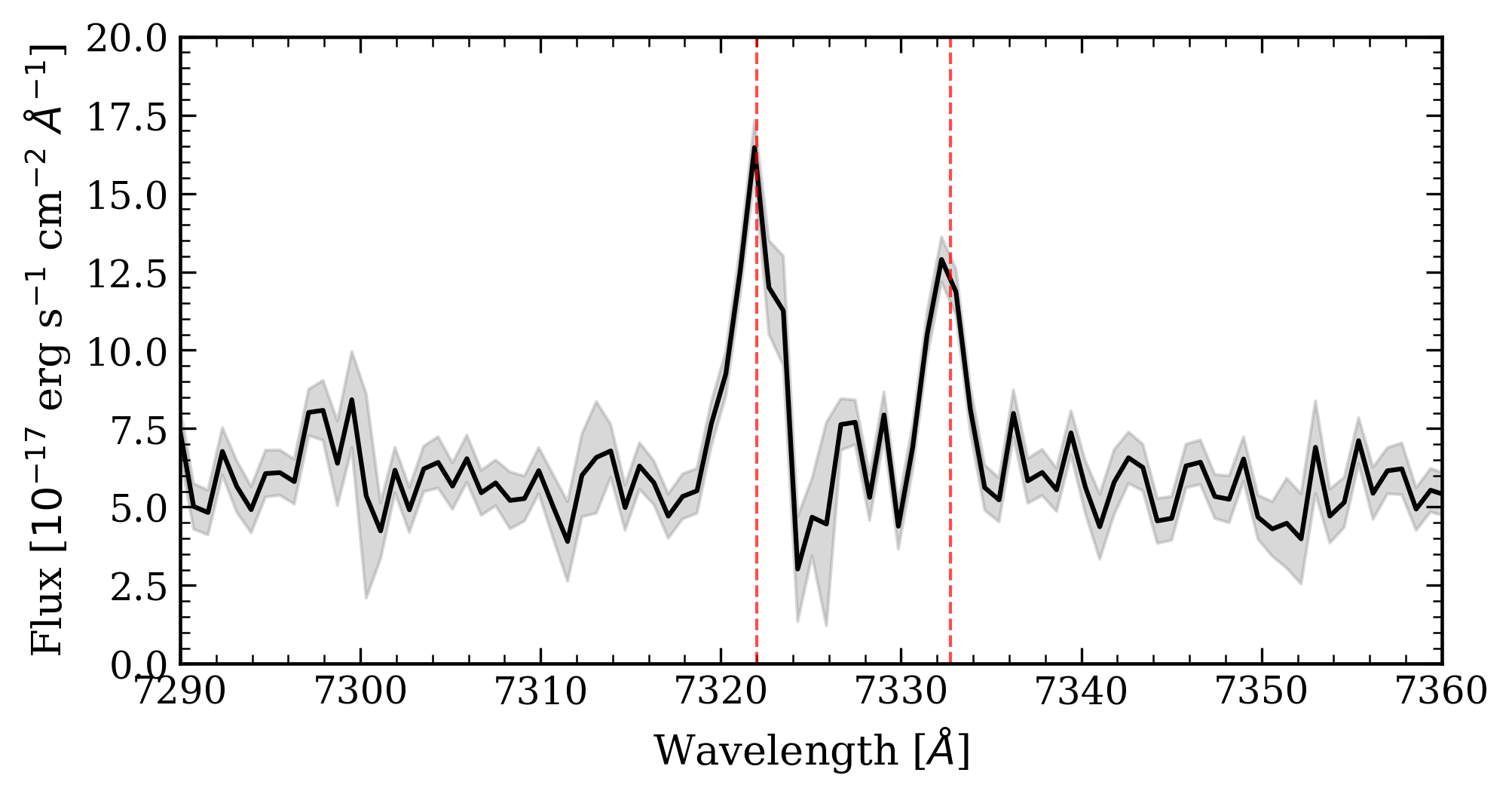}}
\caption{%
    Rest-frame DESI spectrum of I Zw 18 showing clear detection of [\ion{O}{II}]$\lambda\lambda$7320,7330 doublet. Gray shared area represents errorbars. Dashed vertical lines mark position of emission lines.
}
\label{figure:spectrum}
\end{figure}

\begin{figure}
\resizebox{0.97\hsize}{!}{\includegraphics[angle=000]{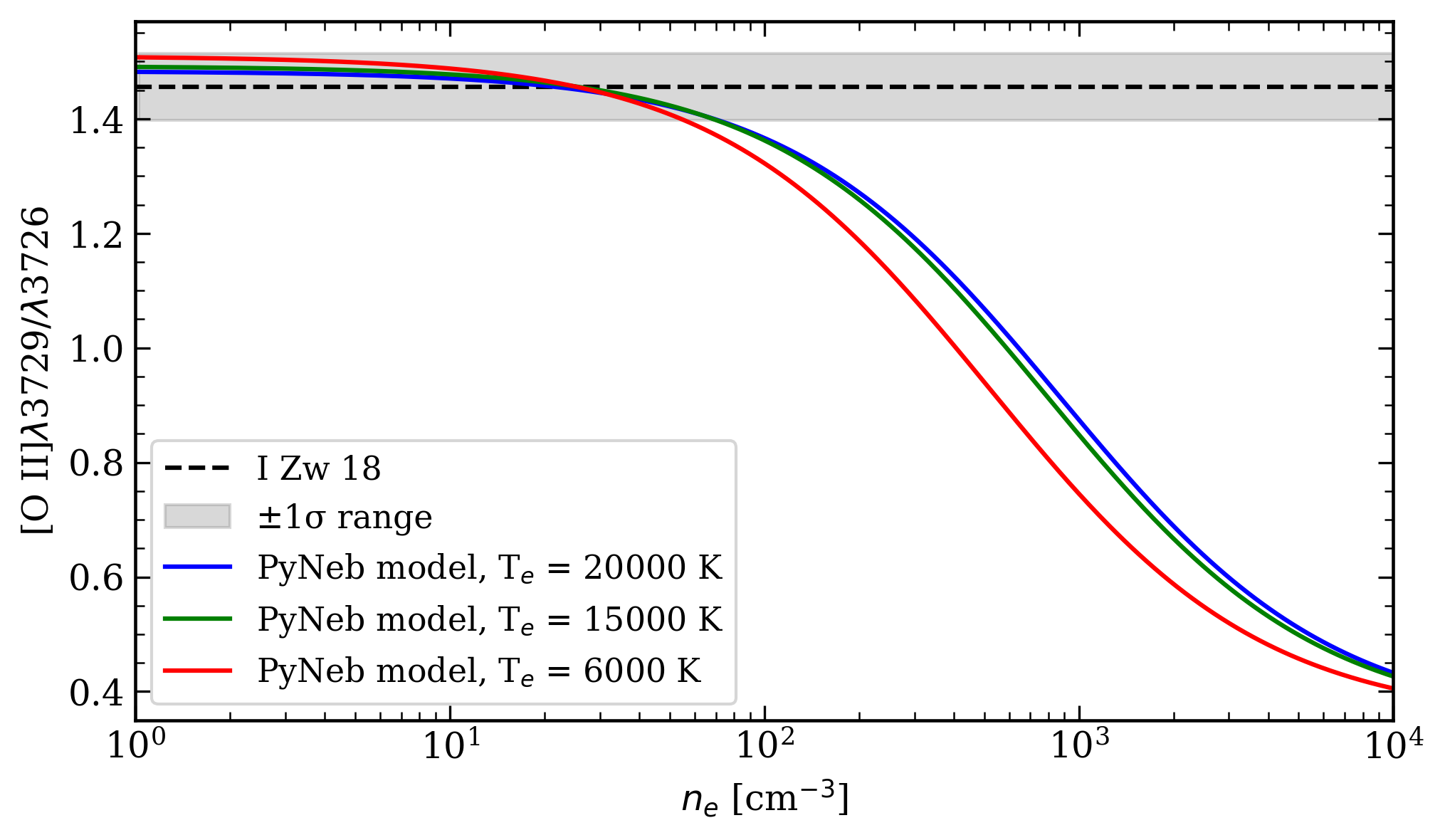}}
\caption{%
    [\ion{O}{II}]$\lambda$3729/[\ion{O}{II}]$\lambda$3726 ratio as a function of n$_e$. Blue solid line represents PyNeb model for T$_e = 20\,000$~K, typical for extremely low metallicity \ion{H}{II}~regions. For comparison, red solid line represents PyNeb model for T$_e = 6\,000$~K. Dotted line and gray area show [\ion{O}{II}]$\lambda$3729/[\ion{O}{II}]$\lambda$3726 ratio derived from DESI spectrum and its uncertainty.
}
\label{figure:density-OII}
\end{figure}

\section{Data}
\label{sect:data}

New spectrum of the SE knot in I Zw 18 has been published in DESI DR1. Fig.\ref{figure:image} shows the exact position of the DESI aperture, which has a diameter of 1.5", which corresponds to 138~pc at 19~Mpc distance~\citep{Fiorentino2010}.
Emission line measurements were obtained from \textsc{FastSpecFit} Spectral Synthesis and Emission-Line Catalog designed for stellar continuum and emission-line modeling tailored for DESI \citep{FastSpecFit2023}. 
It integrates physically-motivated stellar population synthesis and emission-line templates to simultaneously model DESI's optical spectrophotometry from three cameras, alongside ultraviolet to infrared broadband photometry. Along with other data, it provides emission line fluxes and corresponding errors. Table~\ref{table:flux} provides a summary of emission lines and their fluxes used in our analysis. The signal-to-noise ratio in all listed emission lines is $> 5$. Along with [\ion{O}{III}]$\lambda$4363 and [\ion{S}{III}]$\lambda$6312 auroral emission lines, the detection of the auroral [\ion{O}{II}]$\lambda\lambda$7320,7330 lines (see Fig.\ref{figure:spectrum}), makes possible the first direct measurement of the T$_e$([\ion{O}{II}]) in I~Zw~18.

\section{Determination of the chemical abundances, electron densities and temperatures}
\label{sect:abundance}

In this section, we describe our methodology for determining electron densities (n$_e$) and temperatures (T$_e$) along with chemical abundances. The calculations were performed using PyNeb code version 1.1.24~\citep{Luridiana2015} with the PyNeb atomic data dictionary PYNEB\_23\_01.

\subsection{Extinction correction and electron density}

The Balmer line ratios measured in the DESI spectrum of the SE knot show systematic deviations from the theoretical Case~B values, particularly for the higher-order Balmer lines, which cannot be explained by dust extinction alone. A more detailed analysis of the Balmer decrement, including impact of stellar absorption, and suggested flux corrections are presented in Appendix~\ref{appendix:A}.

The spectral resolution of the DESI data enables us to resolve the [\ion{O}{II}]$\lambda\lambda$3727,3729 doublet, as well as the [\ion{S}{II}]$\lambda\lambda$6716,6731 doublet. For I Zw 18 SE knot, we calculated line ratios of [\ion{O}{II}]$\lambda$3729/[\ion{O}{II}]$\lambda$3726 = 1.456$\pm$0.058 and [\ion{S}{II}]$\lambda$6716/[\ion{S}{II}]$\lambda$6731 = 1.473$\pm$0.066. From these diagnostic ratios, we found the electron density n$_e$ from [\ion{O}{II}]$\lambda$3729/[\ion{O}{II}]$\lambda$3726 is equal to 27$\pm$32. The changes in n$_e$ are negligible for T$_e > 10\,000$~K, typical for extremely low metallicity \ion{H}{II}~regions. Fig.\ref{figure:density-OII} presents a comparison between the observed [\ion{O}{II}] doublet flux ratio and theoretical ratios at varying n$_e$. The figure demonstrates that the theoretical relationship between n$_e$ and [\ion{O}{II}]$\lambda$3729/[\ion{O}{II}]$\lambda$3726 does not significantly depend on T$_e$ in the low density regime. [\ion{O}{II}] doublet probes n$_e$ in the O$^+$ ionization zone, for which later we will derive T$_e$ and the oxygen abundance. However, due to substantial uncertainty in n$_e$ estimation at low density, the computed n$_e$ value will not be utilized in determining T$_e$ and chemical abundances. Instead, a fixed value of n$_e = 100$~cm$^{-3}$ will be assumed since its effect is negligible in the calculation of the final ionic abundances at low density~\citep{Osterbrock2006}.
Nonetheless, we additionally checked our conclusion on the low density regime using [\ion{S}{II}]$\lambda$6716/[\ion{S}{II}]$\lambda$6731 ratio and found that both [\ion{S}{II}] and [\ion{O}{II}] doublets consistently show a low-density state (n$_e < 100$~cm$^{-3}$). 

\subsection{Electron temperatures}

From the combination of the nebular [\ion{O}{III}]$\lambda\lambda$4959,5007 and auroral [\ion{O}{III}]$\lambda$4363 lines, we obtained T$_e$([\ion{O}{III}]) = 21\,200$\pm$860~K. This value is higher than $19\,600\pm600$~K reported for the SE knot by \citet{Kehrig2016} and than averaged T$_e = 18\,500\pm2\,000$~K measured across entire galaxy by \citet{RickardsVaught2025}. However, both studies report significant temperature variations ($\sim15\,000$–$24\,000$~K) at the galaxy scales as well as locally within its SE knot. Therefore, slightly elevated, with respect to averaged estimations by previous IFS studies, value obtained from DESI spectrum likely caused by differences in spatial sampling. 
T$_e$(\ion{O}{II}) computed using [\ion{O}{II}]$\lambda\lambda$3729,3726/[\ion{O}{II}]$\lambda\lambda$7320,7330 ratio and representing the lower ionization zone, turned out to be significantly lower, T$_e$([\ion{O}{II}]) = 16\,170$\pm$950~K.

As it has been summarized by \citet{MendezDelgado2023}, measurements of T$_e$([\ion{O}{II}]) can be affected by several effects such as uncertainty in the reddening correction or the quality of the flux calibration, recombination contribution to the CELs, temperature fluctuations, density variations. However, we corrected emission line fluxes to make them consistent with the theoretical Balmer decrement, addressing potential issues with reddening and/or flux calibration. Possible temperature fluctuations within the O$^+$ zone may lead to overestimation of the T$_e$([\ion{O}{II}]). Therefore, the average temperature in the O$^+$ zone can be lower than T$_e$([\ion{O}{II}]) and should be considered as an upper limit. Possible density fluctuations will have similar effect. Both [\ion{O}{II}]$\lambda$3729/[\ion{O}{II}]$\lambda$3726 and [\ion{S}{II}]$\lambda$6716/[\ion{S}{II}]$\lambda$6731 ratios indicate to low density region in both ionization zones. Although small high-density knots cannot be excluded, their contribution can only lower the average T$_e$ since a higher n$_e$ leads to a lower T$_e$ at a given [\ion{O}{II}]$\lambda\lambda$3729,3726/[\ion{O}{II}]$\lambda\lambda$7320,7330 ratio.

On the other hand, direct T$_e$ estimations in low ionization zones at extremely low metallicities are crucial because there are indications of non-linear relation between T$_e$ in zones of high and low ionization. Furthermore, the number of data points with simultaneous measurement of both temperatures in T$_e$([\ion{O}{III}]) > 14\,000~K range is extremely limited \citep[see, e.g.,][]{ArellanoCordova2020,MendezDelgado2023} while establishing a reliable relation between T$_e$([\ion{O}{III}]) and T$_e$([\ion{O}{II}]) or T$_e$([\ion{N}{II}]) is important to accurately determine total oxygen abundance and the ICFs for abundances of other chemical elements, which depend sensitively on O$^+$/O$^{++}$. 

\begin{figure}
\resizebox{0.95\hsize}{!}{\includegraphics[angle=000]{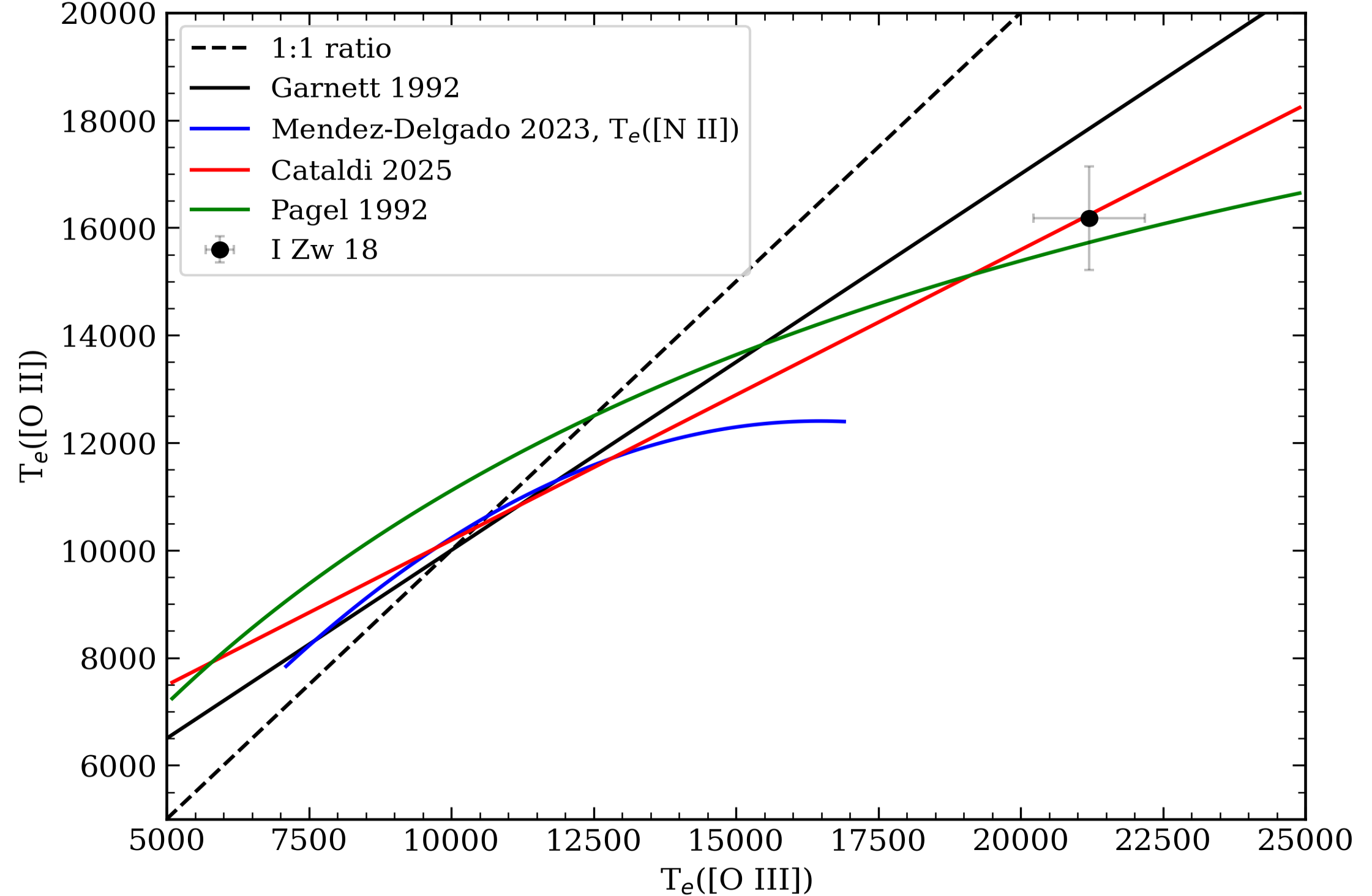}}
\caption{%
    T$_e$([\ion{O}{II}]) as a function of T$_e$([\ion{O}{III}]) for model from \citet{Garnett1992} (solid black line) in comparison with T$_e$([\ion{O}{II}]) and T$_e$([\ion{O}{III}]) derived for SE knot in I Zw 18 (black circle). Solid blue line represents quadratic model from \citet{MendezDelgado2023} for T$_e$([\ion{N}{II}]) as a function of  T$_e$([\ion{O}{III}]), while red and green lines represent relations from \citet{Cataldi2025} and \citet{Pagel1992}, respectively. For this model we preserved limited T$_e$([\ion{O}{III}]) representing the range of T$_e$([\ion{O}{III}]) in its calibration sample.
}
\label{figure:t2t3}
\end{figure}

It is interesting to compare our direct T$_e$([\ion{O}{II}]) estimation with the one obtained using popular \( t_2-t_3 \) relations since they are widely used for calculating the metallicity of metal-poor galaxies when none of auroral lines in low ionization zones are available. Linear \( t_2-t_3 \) relation by \citet{Garnett1992} and our estimation of T$_e$([\ion{O}{III}]) gives T$_e^{rel}$([\ion{O}{II}]) = 17840$\pm$600~K, which is 16\% higher compared to direct estimation using [\ion{O}{II}]$\lambda\lambda$7320,7330 lines. This result differs from the conclusion in \citet{MendezDelgado2023} where they found that T$_e$([\ion{N}{II}]) at high temperature is significantly lower than predicted by linear relations similar to \citet{Garnett1992}. Although the maximum T$_e$([\ion{O}{III}]) in the \citet{MendezDelgado2023} sample is limited to $\sim 14\,000$~K, with our measurement we extend the range up to 21200~K. The measured $T_e([\mathrm{O\,II}])$ in I~Zw~18 is also relevant for high-redshift star-forming galaxies, which typically exhibit high ionization parameters and low metallicity. Some previous studies have shown that the \( t_2-t_3 \) relation may depend on the ionization parameter \citep[e.g.][]{Pilyugin2007,ArellanoCordova2020}. In this context, I~Zw~18 serves as a local analog of high-$z$ systems. As shown in Fig.\ref{figure:t2t3}, our temperature measurements for I~Zw~18 are the most consistent with \( t_2-t_3 \) relations by \citet{Cataldi2025}, developed considering high redshift objects, and \citet{Pagel1992} based on photoionization models. These facts indicate that \( t_2-t_3 \) at high temperatures may be more complex that one dimensional relation and highlight the need of further studies of temperature structure at low metallicities.

Detection of the [\ion{S}{III}]$\lambda$6312 line enabled the calculation of the T$_e$ in the medium-temperature [\ion{S}{III}] zone, T$_e$([\ion{S}{III}]). Since {[\ion{S}{III}]$\lambda$9531}/{[\ion{S}{III}]$\lambda$9069} ratio differs substantially from 2.47, which is the theoretical value. It may indicate telluric absorptions in {[\ion{S}{III}]$\lambda$9069} line. Therefore, we used only {[\ion{S}{III}]$\lambda$9531} nebular line to compute T$_e$([\ion{S}{III}]). We determined T$_e$([\ion{S}{III}]) = 17290$\pm$1750~K, which falls between T$_e$([\ion{O}{II}]) and T$_e$([\ion{O}{III}]).

\subsection{Chemical abundances}

As we derived T$_e$ directly from auroral to nebular line ratios in the [\ion{O}{II}] and [\ion{O}{III}] zones, we can estimate the total oxygen abundance using these T$_e$([\ion{O}{II}]) and T$_e$([\ion{O}{II}]) values together with fluxes of [\ion{O}{II}]$\lambda\lambda$3729,3726 and [\ion{O}{III}]$\lambda\lambda$4959,5007 assuming low-density regime. Negligible ionic O$^{3+}$ abundance is expected since \ion{He}{II}$\lambda4686$ line is not detected in the spectrum. Applying the PyNeb {\it getIonAbundance} method and combining O$^{+}$ and O$^{++}$ abundances, we obtained 12+$\log$(O/H) = 7.066$\pm$0.046.

Since in previous studies neither [\ion{O}{II}] nor [\ion{N}{II}] auroral lines were measured, they relied on \( t_2-t_3 \) relations to estimate O/H. Therefore, we also estimate O/H by computing T$_e$([\ion{O}{II}]) from T$_e$([\ion{O}{III}]) using \( t_2-t_3 \) relation from \citet{Garnett1992}. In this case, we obtained 12+$\log$(O/H)$_{t2t3}$ = 7.036$\pm$0.035, which is compatible with our estimation using direct measurements of both T$_e$([\ion{O}{II}]) and T$_e$([\ion{O}{II}]). The alternative quadratic \( t_2-t_3 \) relation from \citet{MendezDelgado2023} extends only to 17\,000~K and, therefore, cannot be applied to I~Zw~18 without significant extrapolation. Recently, \citet{RickardsVaught2025} applied 3- and 2-state models and parametric \( t_2-t_3 \) relation dependent on excitation parameter and found the oxygen abundance of 7.2$\pm$0.02~dex which is 0.15~dex higher than our estimate.
Thus, different \( t_2-t_3 \) relations provide varying estimates of total O/H, highlighting the need for further studies to better constrain these relations and T$_e$([\ion{O}{II}]) for low metallicity galaxies.

Assuming \( N/O \approx N^+/O^+ \) and T$_e$([\ion{O}{II}]) $\approx$ T$_e$([\ion{N}{II}]), we obtained nitrogen-to-oxygen ratio $\log$(N/O) = -1.509$\pm$0.097, which is consistent with average value for metal-poor galaxies from \citet{Zinchenko2024}. Using methodology described in \citet{Zinchenko2024} we also derived $\log$(S/O) = -1.558$\pm$0.041. This value is very close to an average S/O ratio for metal-poor galaxies as reported in \citet{Zinchenko2024}.

\section{Conclusions}
\label{section:Summary}

This study presents the first direct measurement of electron temperature in the low ionization zone of I~Zw~18, enabled by the detection of the [\ion{O}{II}]$\lambda\lambda$7320,7330 auroral doublet in the DESI DR1 spectrum.
We find a substantial temperature difference between ionization zones, with T$_e$([\ion{O}{III}]) = 21200$\pm$860~K exceeding T$_e$([\ion{O}{II}]) = 16170$\pm$950~K. This T$_e$([\ion{O}{II}]) value lies between \( t_2-t_3 \) relations proposed by \citet{Garnett1992} and \citet{MendezDelgado2023} and provides the first direct T$_e$([\ion{O}{II}]) estimation for very high temperatures, far beyond T$_e$([\ion{O}{III}])~$\sim$~17\,000~K, where empirical \( t_2-t_3 \) are not calibrated.
Using direct temperature measurements for both ionization zones, we derive 12+$\log$(O/H) = 7.066$\pm$0.046. 
We also derived $\log$(N/O) = -1.509$\pm$0.097 and $\log$(S/O) = -1.558$\pm$0.041, which are consistent with average values for metal-poor galaxies from \citet{Zinchenko2024}. 

Our results extend the calibration of T$_e$ relations to the highest temperatures, providing important anchor points for the temperature structure of extremely metal-poor \ion{H}{II} regions. Potential systematic biases in metallicities derived using empirical or theoretical temperature relations may have important implications for studies of metal-poor galaxies (e.g. primordial helium abundance derivation), including high-redshift galaxies, where direct T$_e$ measurements in [\ion{O}{II}] zone are mostly impossible and abundance estimates rely on such relations. Therefore, accurate chemical abundance determinations in primordial environments require direct T$_e$ measurements in all ionization zones or robust \( t_2-t_3 \) relations, highlighting the need for improved models of temperature structures in extremely metal-poor galaxies.

\begin{acknowledgements}
We are grateful to the referee for his or her constructive comments. 
IAZ acknowledges funding from the Deutsche Forschungsgemeinschaft (DFG; German Research Foundation)---project-ID 550945879.
JVM acknowledge financial support from the Spanish MINECO grant PID2022-136598NB-C32 and from the AEI  “Center of Excellence Severo Ochoa” award to the IAA (SEV-2017-0709). PP acknowledges support by Funda\c{c}\~{a}o para a Ci\^{e}ncia e a Tecnologia (FCT) grants UID/FIS/04434/2019, UIDB/04434/2020, UIDP/04434/2020 and Principal Investigator contract CIAAUP-092023-CTTI.
JEM-D thanks the support of the SECIHTI CBF-2025-I-2048 project “Resolving the Internal Physics of Galaxies: From Local Scales to Global Structure with the SDSS-V Local Volume Mapper” (PI: Méndez-Delgado).\\
This research used data obtained with the Dark Energy Spectroscopic Instrument (DESI). DESI construction and operations are managed by the Lawrence Berkeley National Laboratory. This material is based upon work supported by the U.S. Department of Energy, Office of Science, Office of High-Energy Physics, under Contract No. DE–AC02–05CH11231, and by the National Energy Research Scientific Computing Center, a DOE Office of Science User Facility under the same contract. Additional support for DESI was provided by the U.S. National Science Foundation (NSF), Division of Astronomical Sciences under Contract No. AST-0950945 to the NSF’s National Optical-Infrared Astronomy Research Laboratory; the Science and Technology Facilities Council of the United Kingdom; the Gordon and Betty Moore Foundation; the Heising-Simons Foundation; the French Alternative Energies and Atomic Energy Commission (CEA); the National Council of Science and Technology of Mexico (CONACYT); the Ministry of Science and Innovation of Spain (MICINN), and by the DESI Member Institutions: www.desi.lbl.gov/collaborating-institutions.
\end{acknowledgements}

\bibliography{reference}

\begin{appendix}

\section{Emission line fluxes and correction for extinction}
\label{appendix:A}

\begin{table}
\caption{Emission line fluxes normalized to H$\beta = 100$.}
\label{table:flux}      
\centering                          
\begin{tabular}{l r r r c c c c c c c}  
\hline\hline                 
Line & FastSpecFit\tablefootmark{a}  & Gaussian fit\tablefootmark{b} & Case B\tablefootmark{c} & Corr\tablefootmark{d} \\    
\hline                        
 {[\ion{O}{II}]$\lambda$3726}   &  22.71$\pm$0.70 & &  & 0.88 \\ 
 {[\ion{O}{II}]$\lambda$3729}   &  33.07$\pm$0.85 & &  & 0.88 \\
       H$\delta$                &  29.73$\pm$0.54 & 30.47$\pm$1.18 &  26.4 & 0.88 \\
       H$\gamma$                &  52.41$\pm$0.72 & 50.91$\pm$0.99 &  47.5 & 0.9 \\
 {[\ion{O}{III}]$\lambda$4363}  &   6.96$\pm$0.43 & &  & 0.9 \\
       H$\beta$                 & 100.00$\pm$0.65 & 100.00$\pm$1.04 &  100.0 \\
 {[\ion{O}{III}]$\lambda$4959}  &  58.27$\pm$0.49 & &  \\
 {[\ion{O}{III}]$\lambda$5007}  & 169.25$\pm$0.89 & &  \\
 {[\ion{S}{III}]$\lambda$6312}  &   0.67$\pm$0.10 & &  \\
 {[\ion{N}{II}]$\lambda$6548}   &   0.49$\pm$0.09 & &  \\
       H$\alpha$                & 264.90$\pm$0.95 & 278.02$\pm$2.56 &  274.6 \\  
 {[\ion{N}{II}]$\lambda$6584}   &   1.44$\pm$0.09 & &  \\
 {[\ion{S}{II}]$\lambda$6716}   &   3.94$\pm$0.11 & &  \\
 {[\ion{S}{II}]$\lambda$6731}   &   2.67$\pm$0.10 & &  \\
 {[\ion{O}{II}]$\lambda$7320}\tablefootmark{e}   &   0.89$\pm$0.08 & &  \\
 {[\ion{O}{II}]$\lambda$7330}\tablefootmark{f}   &   0.70$\pm$0.06 & &  \\
 {[\ion{S}{III}]$\lambda$9069}  &   3.07$\pm$0.07 & &  \\
 {[\ion{S}{III}]$\lambda$9531}  &   9.43$\pm$0.13 & &  \\
\hline    
\end{tabular}
\tablefoot{
\tablefoottext{a}{Flux from FastSpecFit VAC.}
\tablefoottext{b}{Flux from our Gaussian fit without subtraction of stellar continuum.}
\tablefoottext{c}{Recombination line flux expected from Case~B.}
\tablefoottext{d}{Flux correction coefficient.}
\tablefoottext{e}{The line corresponds to unresolved doublet [\ion{O}{II}]$\lambda\lambda$7319,7320.}
\tablefoottext{f}{The line corresponds to unresolved doublet [\ion{O}{II}]$\lambda\lambda$7329,7330.}
}
\end{table}

\begin{figure}
\resizebox{1.00\hsize}{!}{\includegraphics[angle=000]{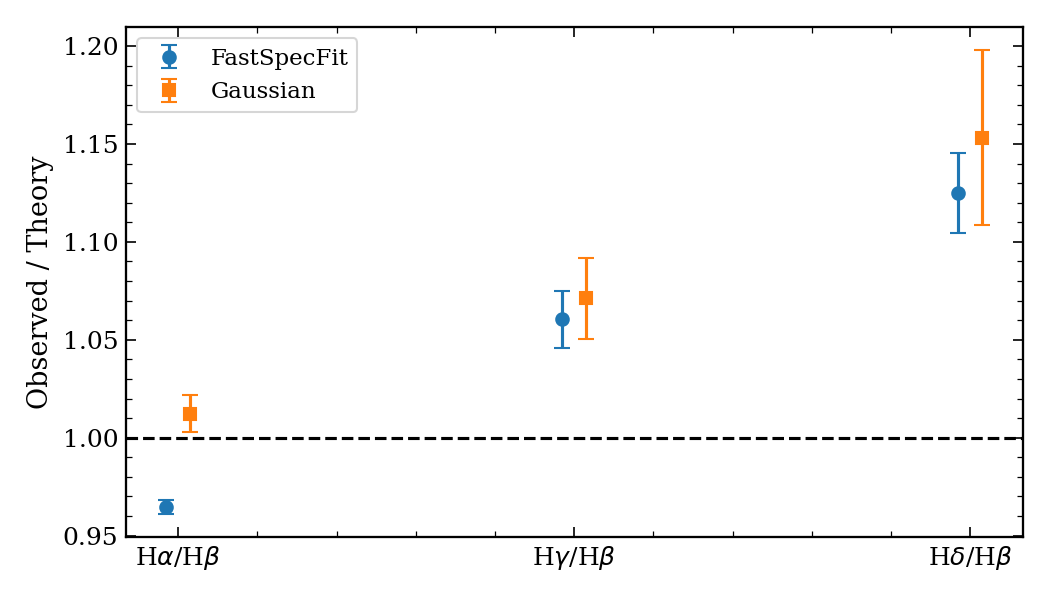}}
\caption{%
    Observed-to-theoretical Balmer line ratios (H$\alpha$/H$\beta$, H$\gamma$/H$\beta$, H$\delta$/H$\beta$) for two measurement methods. Blue circles represent full spectral fitting including stellar absorption component from \textsc{FastSpecFit} catalog. Orange squares represent our fitting with single Gaussian without subtraction of stellar continuum. Error bars show $1\sigma$ uncertainties, and the dashed line marks the consistency with Case~B recombination at n$_e = 100$~cm$^{-3}$ and T$_e = 20\,000$~K.
}
\label{figure:balmer}
\end{figure}

\begin{figure}
\resizebox{1.00\hsize}{!}{\includegraphics[angle=000]{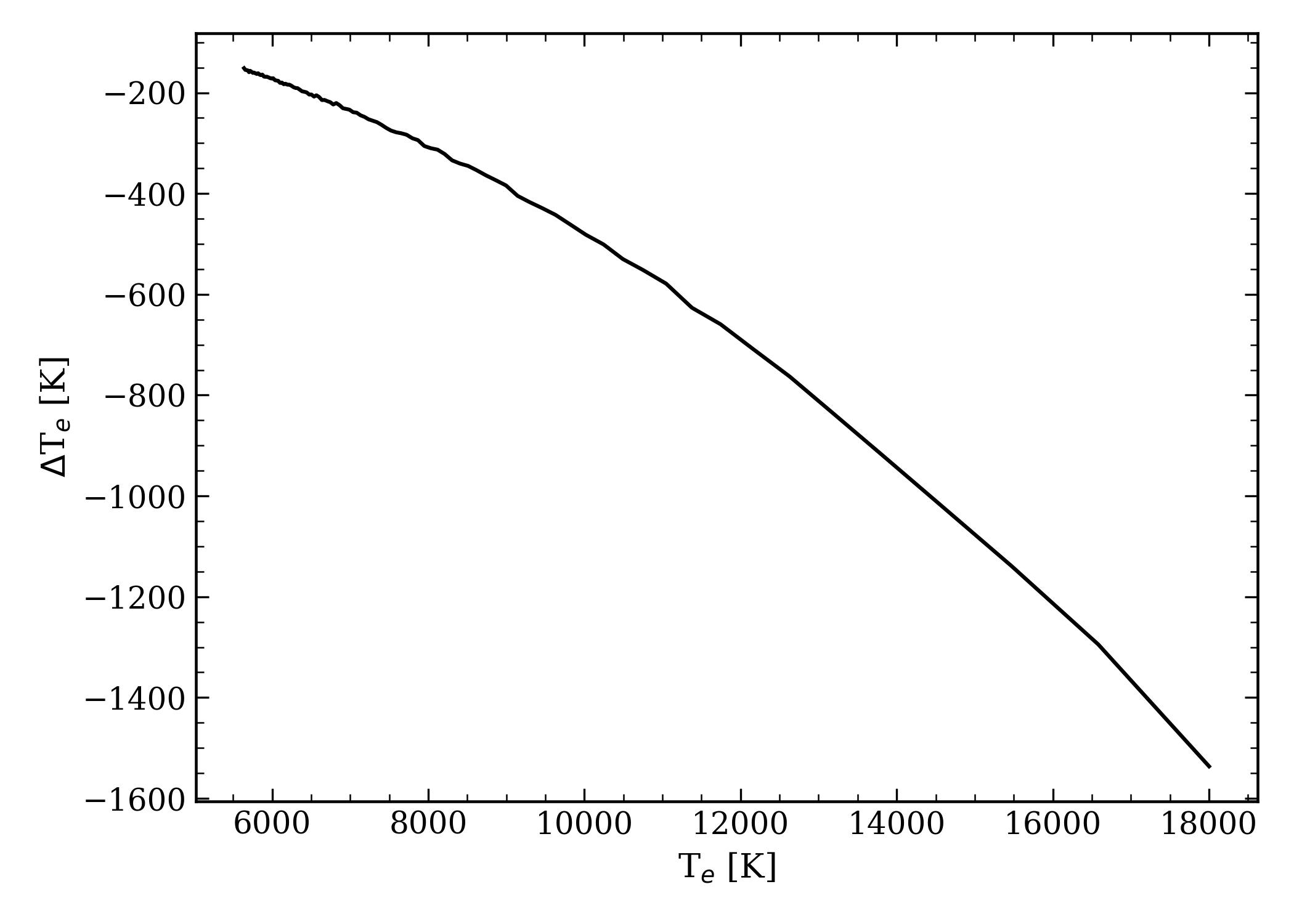}}
\caption{%
    Difference in the derived T$_e$([O II]) caused by a 10\% increase in the [\ion{O}{II}]$\lambda\lambda$3726,3729 flux as a function of the original temperature.
}
\label{figure:calibration-model}
\end{figure}

We summarize emission line measurements from the VAC \textsc{FastSpecFit} catalog in Table~\ref{table:flux}. The fluxes returned by \textsc{FastSpecFit} are already corrected for Galactic extinction. This table shows that the measured H$\alpha$/H$\beta$ ratio ($2.649 \pm 0.020$) is $\simeq3\%$ lower than the theoretical value expected at $T_e = 20\,000$~K, formally yielding a negative $c(\mathrm{H}\beta) = -0.045 \pm 0.009$ when applying the extinction curve of \citet{Fitzpatrick1999}, while previous studies reported a small positive $c(\text{H}\beta)$ values  \citep{Skillman1993,Vilchez1998,Izotov1994,Kehrig2016}. \citet{Ucci2019} using the same IFU data as \citet{Kehrig2016}, also reported very low but positive visual extinction values, A$_V \sim 0.01$~mag, in most spaxels. 

At the same time, the observed H$\gamma$/H$\beta$ and H$\delta$/H$\beta$ ratios exceed their theoretical values by $\sim6$--12\%, also implying negative $c(\mathrm{H}\beta)$. For the physical conditions typical of metal-poor \ion{H}{ii} regions, we adopt Case~B Balmer ratios at n$_e = 100$~cm$^{-3}$ and T$_e = 20\,000$~K computed with \textsc{PyNeb}. The theoretical ratios used in this work are listed in Table~\ref{table:flux}. These Balmer ratios therefore cannot be explained by a physically plausible dust extinction correction alone.

To test whether overestimation of absorption features during stellar continuum modelling can explain these discrepancies, we refit all Balmer emission lines using single-Gaussian profiles without subtracting stellar component. While this brings H$\alpha$/H$\beta$ closer to the theoretical ratio, it does not significantly change H$\gamma$/H$\beta$ or H$\delta$/H$\beta$, which remain discrepant beyond their uncertainties as shown in Figure~\ref{figure:balmer}. Thus, we conclude that the elevated higher-order Balmer ratios are not caused by the stellar population fitting performed by \textsc{FastSpecFit}.

Thus, we consider a relative flux calibration bias in the blue part of the DESI spectrum as a likely scenario. Although, we cannot rule out possible deviations from the standard Case~B assumptions at very high $T_e$ \citep[e.g.][]{Flury2022,Scarlata2024}. We therefore apply an empirical correction to the blue spectral region, derived from the ratios of the observed to theoretical H$\gamma$/H$\beta$ and H$\delta$/H$\beta$ values. The resulting correction factors correspond to a reduction of the measured fluxes by $\sim12\%$ for [\ion{O}{ii}]$\lambda\lambda3726,3729$ and $\sim10\%$ for [\ion{O}{iii}]$\lambda4363$, based on their proximity in wavelength to the relevant Balmer lines.

Applying this correction increases the derived $T_e(\mathrm{O\,II})$, but lowers $T_e(\mathrm{O\,III})$, bringing our $T_e(\mathrm{O\,III})$ estimate for the SE knot into closer agreement with previous IFU-based studies \citep[e.g.][]{Kehrig2016,RickardsVaught2025}. Increase of [\ion{O}{ii}]$\lambda\lambda3726,3729$ flux by $10\%$ in $T_e(\mathrm{O\,II})$ range of 14\,000--16\,000~K results in $T_e(\mathrm{O\,II})$ uncertainty of 1\,000--1\,200~K, which is compatible with statistical uncertainty of our $T_e(\mathrm{O\,II})$ estimation. Figure~\ref{figure:calibration-model} demonstrates the effect of a systematic increase in the [\ion{O}{ii}]$\lambda\lambda3726,3729$ flux on the derived electron temperature. In this toy model the T$_e$([O II]) is computed using PyNeb over a range of input ratios [\ion{O}{ii}] nebular-to-auroral line ratios (25–250). The x-axis shows the T$_e$ for the reference case, while the y-axis shows the temperature difference $\Delta$T$_e$ between the reference case and a 10\% increase in [\ion{O}{ii}]$\lambda\lambda3726,3729$ flux.

\end{appendix}
 
\end{document}